\documentclass[aps, pra, amsmath, amssymb, reprint, superscriptaddress
]{revtex4-2}
\usepackage{graphicx}
\usepackage{dcolumn}
\usepackage{bm}
\usepackage[mathlines]{lineno}
\usepackage[utf8]{inputenc}
\usepackage[T1]{fontenc}
\usepackage{mathptmx}
\usepackage{etoolbox}
\usepackage{float}
\usepackage{xcolor}

\usepackage[defaultcolor=blue]{changes}

\makeatletter
\def\@email#1#2{%
 \endgroup
 \patchcmd{\titleblock@produce}
  {\frontmatter@RRAPformat}
  {\frontmatter@RRAPformat{\produce@RRAP{*#1\href{mailto:#2}
  {#2}}}\frontmatter@RRAPformat}
  {}{}
}%
\makeatother

\begin{document}

%\preprint{AIP/123-QED}

\title{Uniaxial Magnetic Anisotropy and Type-X/Y Current-Induced Magnetization Switching in Oblique-Angle-Deposited Ta/CoFeB/Pt and W/CoFeB/Pt Heterostructures}
\author{Amir Khan}
\affiliation{New Materials Electronics, Electrical Engineering and Information Technology, Technical University of Darmstadt, Merckstr. 25, 64283 Darmstadt, Germany}
\email{amir.khan@tu-darmstadt.de}
\author{Shalini Sharma}
\affiliation{Advanced Thin Film Technology, Institute of Materials Science, Technical University of Darmstadt, Peter-Grünberg-Straße 2, 64287 Darmstadt, Germany}
\author{Tiago de Oliveira Schneider}
\affiliation{New Materials Electronics, Electrical Engineering and Information Technology, Technical University of Darmstadt, Merckstr. 25, 64283 Darmstadt, Germany}
\author{Markus Meinert}
\affiliation{New Materials Electronics, Electrical Engineering and Information Technology, Technical University of Darmstadt, Merckstr. 25, 64283 Darmstadt, Germany}
\email{markus.meinert@tu-darmstadt.de}

\date {\today}

\begin{abstract}
Planar current-induced magnetization switching (CIMS) driven by spin-orbit torque (SOT) requires an in-plane uniaxial magnetic anisotropy (UMA), which can be induced by oblique-angle sputter deposition of the heavy-metal underlayer in heavy-metal/ferromagnet heterostructures. To enhance the SOT efficiency, we employ trilayer heterostructures of (Ta or W)/CoFeB/Pt, where the CoFeB layer exhibits a UMA of 50\,mT at 2\,nm thickness of Ta or W. The magnetization reversal in Hall-bar devices is detected through unidirectional spin Hall magnetoresistance (USMR) for the type Y geometry (easy-axis transverse to current) and planar Hall measurements for the type X geometry (easy-axis parallel to current). Both configurations exhibit CIMS with sub-microsecond current pulses, reaching switching current densities as low as $2 \times 10^{11}$\,A/m$^2$ for a W (4\,nm)/CoFeB (1.4\,nm)/Pt (2\,nm) stack with a UMA of 146\,mT. Macrospin simulations reproduce the type Y switching as coherent magnetization rotation, whereas the type X devices switch at much lower currents than predicted, indicating that nucleation and domain-wall propagation dominate reversal in this geometry. Our results show that combining oblique-angle deposition with easy-axis engineering enables deterministic, field-free switching, paving the way for future low-power spintronic devices.
\end{abstract}

\maketitle

\section{\label{sec:level1}INTRODUCTION\\}
Electric current-induced spin-orbit torque (SOT) in high-Z metals such as W, Ta and Pt with strong spin-orbit coupling (SOC) has recently\cite{Liu2012} garnered attention in vertical heterostructures for efficiently manipulating the magnetization with low power consumption and high switching speeds. This enables compact and fast memory, and logic gate devices\cite{Manchon2019,Brataas2012,Liu2012}. 
Current magnetic random-access memory (MRAM) uses spin transfer torque (STT)\cite{RALPH20081190} as a method to manipulate the magnetization of nanomagnets. However, STT has some disadvantages, such as the high current density degrading the tunnel barrier during the writing process in magnetic tunnel junctions (MTJs). This issue is not present in SOT-MRAM due to its separate read and write current paths in a three-terminal bit cell, which enhances endurance of the device and overcomes the reliability issues of STT. 

In SOT related experiments, the commonly used sample structure consists of a non-magnetic heavy metal (HM) / ferromagnetic metal (FM) bilayer\cite{Pai2012,Liu2012}, where the HM serves as the source of spin current via the spin Hall effect (SHE)\cite{Sinova2015}, and the FM layer acts as the spin current sink. The spin current excites magnetization dynamics as described by the Landau-Lifshitz-Gilbert-Slonczewski (LLGS) equation\cite{Slonczewski1996}. Depending on the magnetic anisotropy of the FM layer, SOT-driven switching can be realized in systems with either in-plane magnetic anisotropy (IMA)\cite{Liu2020} or perpendicular magnetic anisotropy (PMA)\cite{You2015}. The HM/FM interface acts as a source of SOT that arises from spin polarization generated by SHE and the Rashba effect\cite{Sinova2015,Obata2008} in the HM underlayer with high SOC.

In conventional HM/FM bilayers, the SOT efficiency can be limited by use of a single HM layer as the source of spin current. To overcome this limitation in recent studies\cite{Huang2018,Zheng2019,Skowroski2019}, it is demonstrated theoretically and experimentally that if an ultrathin FM layer is sandwiched between two HM layers, which leads to a trilayer structure HM1/FM/HM2 where top and bottom HM1 and HM2 layers must have opposite spin Hall angles (SHA) such as Ta and Pt. The two HM layers enable spin current injection from both interfaces with identical polarization and enhance the SOT efficiency relative to the bilayer system. For example, the SOT efficiency of a W/CoFeB/Pt stack was enhanced by a factor of 1.5 compared to the bilayer W/CoFeB\cite{Skowroski2019,Huang2018}.
\begin{figure*}
\includegraphics{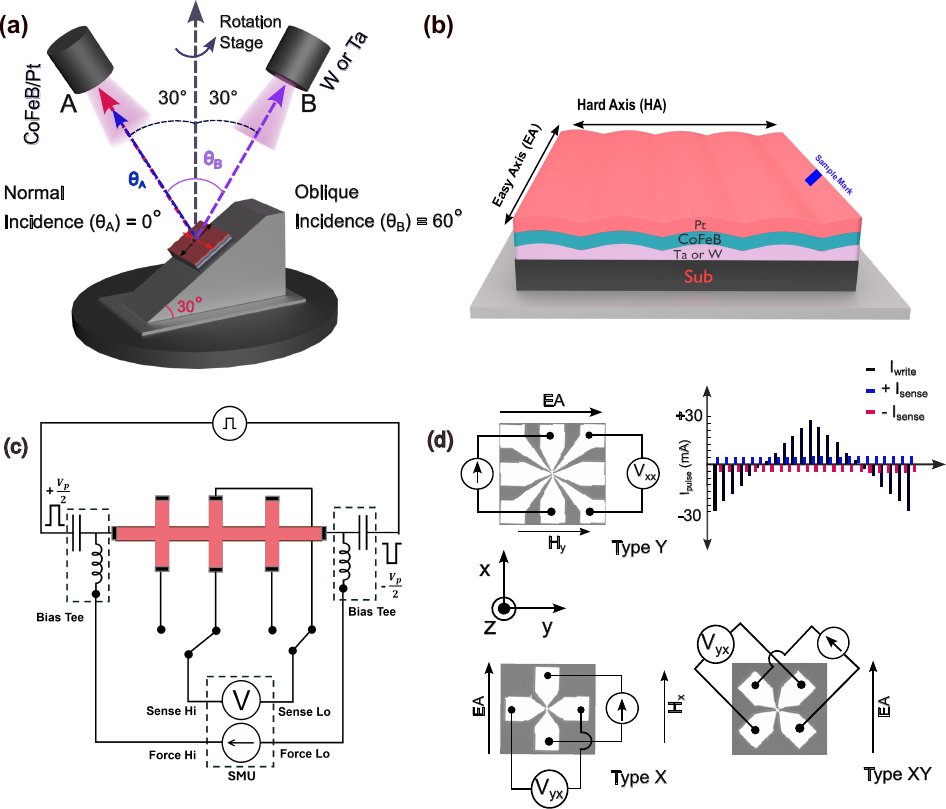}
 \caption{\label{fig : setup}  Sample deposition geometry and CIMS measurement setup. (a) W/Ta underlayer deposited at $\theta_\mathrm{B} =  60^{\circ}$ and CoFeB/Pt at $\theta_\mathrm{A} =  0^{\circ}$. (b) Samples were marked along the HA direction. The EA is parallel to the microscopic ripple structure\cite{McMichael2000}. (c) In the Hall cross, write current is injected with opposite polarities ($\pm V_P/2$), generating a virtual ground at the device center. This configuration effectively avoids current shunting into the transverse voltage probes and enforces write current propagation along the longitudinal current channel. (d) Schematic illustration of type Y, type X and type XY switching scheme with current channel width $5\,\mu \text{m}$. In-plane magnetic field $H_{\mathrm{y}}$ is applied transverse to the current channel in type Y, $H_{\mathrm{x}}$ is applied parallel to current channel in type X, and at  $45^{\circ}$ in type XY. The measurement sequence of $\Delta R_{\mathrm{xx}}$ and $\Delta R_{\mathrm{yx}}$ is also shown.}
\end{figure*}

SOT-driven magnetization switching occurs in mainly three configurations: type X, type Y and type Z. If the FM layer has IMA with applied current $I$ parallel or transverse to the easy axis (EA) then it is called type X ($\mathrm{EA} \parallel I$) or type Y ($\mathrm{EA} \perp I$) but if the FM layer has PMA, then it is called type Z. In the type X geometry, if the current channel is canted from the EA by a few degrees then it is called type XY\cite{Hu2025,Khang2020}. In a typical HM/FM heterostructure, the spin current with polarization $(\bm{\sigma})$ generates a damping-like ($\tau_\mathrm{DL} \propto B_\mathrm{DL} \cdot \left(\bm{m} \times \left(\bm{\sigma} \times \bm{m}\right)\right)$) and a field-like ($\tau_\mathrm{FL} \propto B_\mathrm{FL} \cdot \left(\bm{m} \times \bm{\sigma}\right)$) torque on the magnetization, where $\bm{m}$ is magnetization unit vector and $B_\mathrm{DL/FL} = \frac{\hbar}{2e} \frac{\theta_\mathrm{DL/FL}}{M_\mathrm{s} t_\mathrm{FM}} j$ is the effective field given by the DL and FL SHAs $\theta_\mathrm{DL/FL}$, the saturation magnetization $M_\mathrm{s}$ and the FM thickness $t_\mathrm{FM}$. The DL torque is strong enough to enable current-induced magnetization switching (CIMS) in structures with both in-plane\cite{Liu2012,Fukami2016} and out-of-plane magnetic anisotropy\cite{Miron2011,Liu2012PMA}. However, in type X and type Z configurations, the SOT symmetry needs to be broken by an external out-of-plane field for type X\cite{Posti2023} or an in-plane field parallel to the current for type Z\cite{Torrejon2015}. In contrast, type Y configuration intrinsically have asymmetric torque due to the spin polarization being either parallel or antiparallel to the magnetization\cite{Liu2020,Liu2021}, allowing deterministic switching via direct torque minimization.

Field-free deterministic switching in type Z configuration has nonetheless been widely demonstrated using alternative approaches, including engineering tilted magnetic anisotropy\cite{Kong2019,You2015}, out-of-plane spin polarization\cite{Liu2023TaIrTe4}, domain wall pinning\cite{lee2018field}, unconventional SOT in two-dimensional van der Waals ferromagnets\cite{pandey2025}, SOT-induced exchange bias in antiferromagnet/ferromagnet heterostructures\cite{lin2025}, interlayer Dzyaloshinskii-Moriya interaction\cite{Li2023,Husain2024}, and geometric current bending in perpendicular SOT magnetic tunnel junctions\cite{kateel2023}. More recently it has been reported that a second FM layer with IMA as the bottom layer can provide spin current with both in-plane and out-of-plane components, thereby enabling field-free switching at sub-ns timescales\cite{Yang2024}. However, implementing these approaches in nanoscale devices while maintaining strong PMA remains challenging.

Deterministic field-free magnetization switching in the type-X configuration has been achieved by introducing a slight canting between the easy axis and the current channel, enabling inversion symmetry breaking \cite{Liu2021,Liu2022}. For FM layers with IMA, magnetization detection poses significant challenges and necessitates sophisticated measurement techniques. These include differential planar Hall effect (DPHE) measurements\cite{Mihajlovi2016,Xue2023}, tunneling magnetoresistance (TMR) measurements in three-terminal devices, and non-electrical probing methods such as magneto-optical Kerr effect (MOKE) imaging\cite{shi2019all,wang2019m}. While these established SOT characterization methods require complex experimental protocols (such as DPHE) or challenging fabrication requirements (for TMR devices), recent advances have introduced simpler alternatives: direct current (DC) unidirectional spin Hall magnetoresistance (USMR)\cite{Liu2020} and planar Hall effect (PHE)\cite{Fan2013,Liu2022} techniques as viable readout mechanism for SOT characterization of the HM/FM bilayer system with uniaxial magnetic anisotropy (UMA). 

Methods for achieving UMA with preferred EA include magnetic post-annealing\cite{takahashi1960} and applying a magnetic field during the deposition\cite{kedia2024}. However these methods have limitations when strong and thermally stable magnetic anisotropy is required. Post annealing techniques can cause structural modifications or stress relaxation \cite{Dijken2001,Ono1993} during annealing. Oblique-angle deposition (OAD) is a promising alternative method to enhance the UMA through the formation of ripple structures, as previously reported\cite{Scheibler2023,Fukuma2009,McMichael2000}.

In this manuscript, we report on the current-induced in-plane magnetization switching in trilayer $\mathrm{(Ta\ or\ W)/Co_{40}Fe_{40}B_{20}/Pt}$ structures. We first establish in-plane UMA in $\mathrm{Co_{40}Fe_{40}B_{20}}$ via ripple structures by OAD of the underlayer (W or Ta) enabled by self shadowing\cite{zhu2012,ali2021,soh2014} and adatom steering mechanisms\cite{Dijken1999}. Subsequently, we employ DC USMR and PHE measurements performed on the same sample without the external magnetic field to realize complete current induced magnetization switching in type Y, type X and type XY using sub-microsecond current pulses ($\mathrm{0.4-1\,\mu\text{s}}$). This switching duration is substantially shorter than the millisecond to second pulse width for HM/FM bilayer structures with in-plane magnetization typically reported in the literature\cite{Liu2020,Liu2022,Khang2020}.

\section{\label{sec:level2}METHODS\\}
\subsection{\label{sec:levelA}Sample preparation\\}
A series of samples was deposited at room temperature on $\mathrm {Si/SiO_{2}}$ substrates using a custom magnetron co-sputter deposition system with eight $2\text{"}$ sources by Bestec GmbH, Berlin.  Each sample consisted of $\mathrm{Ta (2\,nm)/Co_{40}Fe_{40}B_{20} (t_\mathrm{FM})/Pt (2\,nm)}$, where $\mathrm{t_\mathrm{FM} = 1.4, 1.6, 1.8 \text{ and} \text{ 2\,nm}}$. In addition, a sample of $\mathrm{W (4\,nm)/Co_{40}Fe_{40}B_{20} (1.4\,nm)/Pt (2\,nm)}$ was deposited to achieve higher SOT torques and anisotropy. Throughout the text, samples names are referred to as Ta/CFB(t$_\mathrm{FM}$)/Pt or W(4)/CFB(1.4)/Pt for brevity. The angle of incidence of the particle beams relative to the substrate carrier normal was $30^{\circ}$, and the substrate to target distance was approximately $\mathrm{\text{12\,cm}}$. The base pressure was lower than $5 \times 10^{-8}$\,mbar and the Ar working pressure was maintained at $\mathrm{2\times{10^{-3}}}\text{\,mbar}$ during the thin film growth. At a $30^{\circ}$ incidence angle, the growth rates were $\mathrm{\text{0.116\,nm/s}}$ for Ta, $\mathrm{\text{0.064\,nm/s}}$ for W, $\mathrm{\text{0.039\,nm/s}}$ for $\mathrm{Co_{40}Fe_{40}B_{20}}$, and $\mathrm{\text{0.111\,nm/s}}$ for Pt. The deposition powers were set to 50\,W in all cases, except for Ta, which was deposited at 100\,W.

The Layer thicknesses were measured by X-ray reflectivity (XRR) and deposition times were calibrated accordingly. For all samples, the underlayer (Ta or W) was deposited at $60^{\circ}$ by mounting the substrate on a $30^{\circ}$ wedge holder and orienting the rotational stage so that the substrate normal was tilted $60^{\circ}$ relative to the particle flux, while CoFeB and Pt layers were deposited at $0^{\circ}$ by rotating the stage to align the substrate normal with the particle flux, with substrate rotation disabled during each deposition (as shown in Fig.~\ref{fig : setup} (a)). No external magnetic field was applied during growth and no post-annealing was performed. After deposition, samples were marked along the hard-axis (HA) direction (as shown in Fig.~\ref{fig : setup} (b)). The HA and easy-axis (EA) directions were subsequently confirmed via longitudinal MOKE (L-MOKE) and anisotropic magnetoresistance (AMR) measurements, as described in a later section. The samples were then patterned into micrometer-scale Hall cross bars with a $\mathrm{5\,\mu\text{m}}$ current channel width using single-step photolithography and ion-beam etching. Three types of current channels were fabricated, oriented at $90^{\circ}$, $45^{\circ}$ and $0^{\circ}$ with the EA on the same sample (as shown in Fig.~\ref{fig : setup} (d)), designated as type Y, type XY and type X respectively.

\subsection{\label{sec:levelB}Measurement setup\\}
We employed the 4-wire probe method to examine longitudinal and transverse magnetoresistance, illustrated in Fig.~\ref{fig : setup} (c). Write-current pulses of identical amplitude but opposite polarity were applied to the longitudinal channel using a custom-built prober station with four $\mathrm{5\,\mu\text{m}}$ tip radius tungsten needles. A Keysight 33500B dual channel arbitrary waveform generator was used to apply synchronized pulses with $180^{\circ}$ phase flip into a Pendulum Instruments F10AD dual channel high-voltage amplifier. Bias tees specified for frequencies from 0.025\,MHz to 100\,MHz were used to separate the pulse path from the DC current path. Rectangular pulses with widths between 400\,ns (100\,ns rise and fall times) and 1000\,ns (250\,ns rise and fall times) were found to be ideally transmitted with the bias tees by monitoring the current on a shunt resistor.

A Keithley 2450 source measure unit (SMU) supplied the DC sense current and measured the corresponding voltage (as shown in Fig.~\ref{fig : setup} (c)). The field scans used for DC USMR and PHE measurements were recorded with alternating polarity of sense current after application of the magnetic field. The write current loops for magnetization switching experiments were recorded according to the measurement sequence shown in Fig.~\ref{fig : setup}(d). After each written current pulse, positive and negative sense currents were used to determine a differential readout of nonlinear effects. This will be discussed in the next section. All current densities are given as average values over the full stack thickness if not stated otherwise.
 
\subsection{\label{sec:levelC}Magnetic anisotropy and electrical detection of magnetization\\}
\subsubsection{\label{sec:levelC1}Uniaxial magnetic anisotropy\\}
\begin{figure}
\includegraphics[width=\columnwidth]{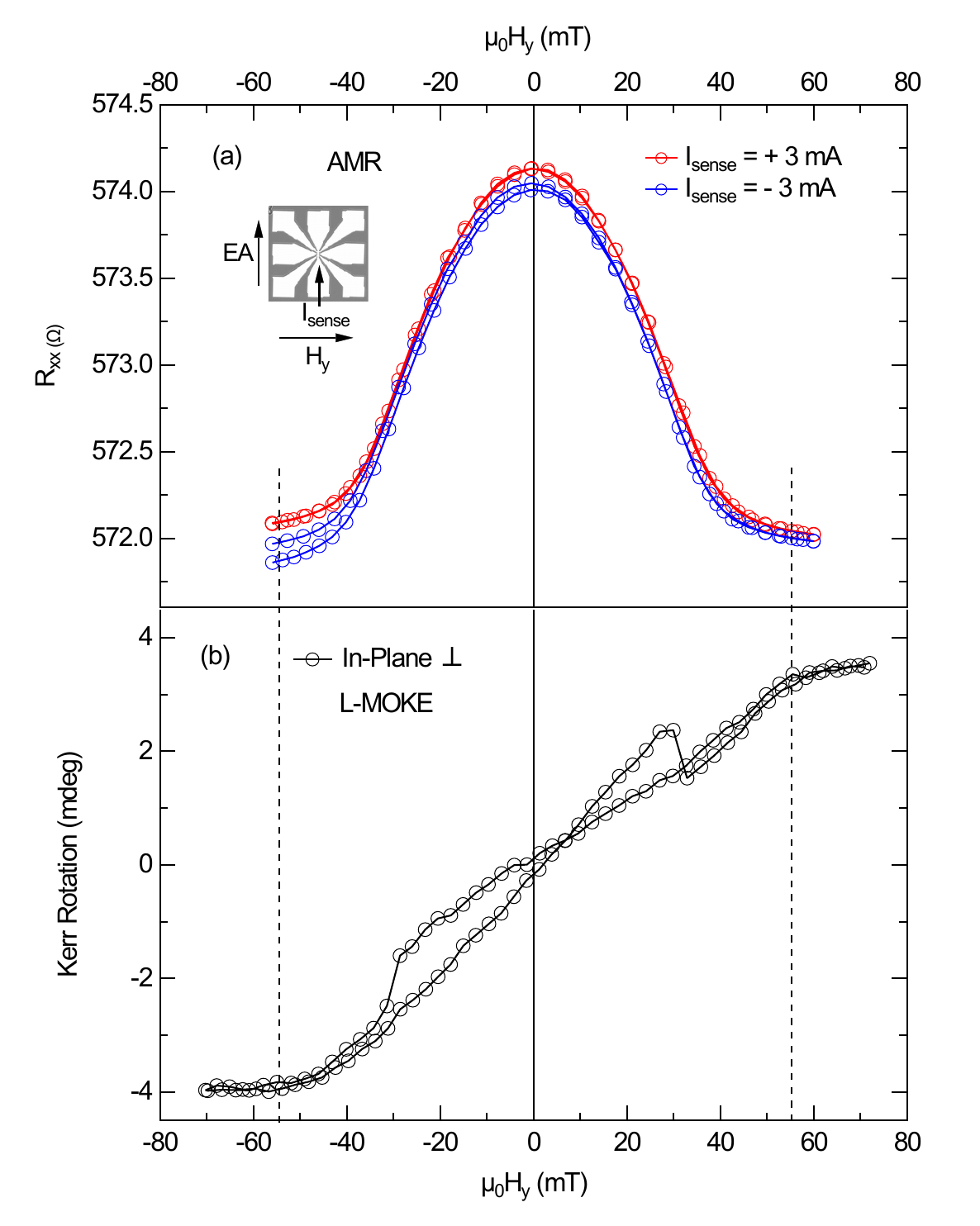}
 \caption{\label{fig : moke}  AMR and L-MOKE magnetic hysteresis loops measurement. (a) Field sweep AMR measurements for device Ta/CFB(2)/Pt by applying the sense current ($I_{\mathrm{sense}} = \pm 3\text{\,mA}$) transverse to the in-plane applied field ($I_{\mathrm{sense}}\parallel \text{EA}$). (b) Magnetic hysteresis loop recorded using L-MOKE while applying the in-plane external magnetic field along the hard axis. The residual hysteresis indicates imperfect alignment of the sample in the L-MOKE setup.}
\end{figure}

The OAD technique is used to induce the UMA with well-defined EA in the FM layer. This can be achieved either by depositing the FM layer at oblique incidence or by growing it on an underlayer deposited at an oblique angle\cite{Fukuma2009}. Among these, the latter approach has been shown to induce more thermally stable anisotropy, since the former shows a reduction in UMA due to microstructural deformation upon heat treatment\cite{Dijken2001,Ono1993}. The magnitude of UMA is strongly dependent on the deposition angle relative to the substrate normal. In our previous work\cite{khan2025}, we demonstrated that increasing the underlayer ($\mathrm{\text{2\,nm}}$ Ta or W) deposition angle enhances UMA, with values approaching ${\mu_{\mathrm{0}}H_{\mathrm{a}}\approx\text{50\,mT}}$ when the angle reached $\theta =  60^{\circ}$. The samples were marked along the particle beam direction, which was subsequently identified as the HA by L-MOKE measurements. Further confirmation was obtained through AMR measurements performed on micrometer-scale Hall crosses. When the in-plane applied field ($H_{\mathrm{y}}$) is along the hard axis, while current is along the EA produces longitudinal resistance variations consistent with an anisotropy field of approximately $\mathrm{\text{50\,mT}}$, as summarized in Fig.~\ref{fig : moke} (a), and (b).

\subsubsection{\label{sec:levelC1}Unidirectional spin Hall magnetoresistance\\}
The USMR effect in HM/FM bilayer systems mimics current-in-plane giant magnetoresistance (CIP-GMR) characteristics and originates from spin accumulation at the HM/FM interface. When current is injected along the x-axis, transverse to the EA (as shown in Fig.~\ref{fig : setup} (d)), SHE or Rashba effects generate spin accumulation. This spin accumulation vector creates a high-resistance state when parallel to the FM magnetization and a low-resistance state when antiparallel. Thus, a small longitudinal resistance change is observed upon magnetization reversal. The origins of the USMR were proposed with multiple mechanisms such as spin-dependent and spin-flip unidirectional magnetoresistance\cite{Avci2018,Chang2021}. The USMR is a promising readout mechanism for magnetization states switched by SOT, and is  particularly effective in type Y configuration where the EA is transverse to the current and spin polarization aligns with the y-axis. The USMR is  a nonlinear transport effect\cite{Avci2015} where $R_{\mathrm{xx}}$ differs for opposite sense current polarities due to the reversal of the interfacial spin accumulation vector. The resistance asymmetry, ${\Delta R_{\mathrm{xx}} = R_{\mathrm{xx}}(+I_{\mathrm{sense}})-R_{\mathrm{xx}}(-I_{\mathrm{sense}})}$, recorded this response.

\subsubsection{\label{sec:levelC1}DC planar Hall effect\\}
The type X configuration has the EA and current along the x-axis while spin polarization $(\bm{\sigma})$ is oriented along the y-axis. Due to the symmetry with $\mathrm{EA} \perp \bm{\sigma}$, there is no USMR response to magnetization reversal, which complicates the detection of deterministic CIMS. In HM/FM bilayers, three mechanisms give rise to transverse voltages with odd parity with respect to the magnetization orientation along the x-axis: a) field-like (FL) torque and planar Hall effect (PHE); b) damping-like (DL) torque and anomalous Hall effect (AHE); c) Joule heating and anomalous Nernst effect (ANE). While the in-plane reorientation due to FL torque is relatively large, the electric response via PHE is small in CoFeB. In contrast, the out-of-plane tilt due to the DL torque is small but the AHE response is large. Thus, both contributions turn out to be of similar magnitude in typical bi- and trilayer systems\cite{Fan2013,Posti2023,Mihajlovi2016}.

The ANE contribution depends strongly on the details of the stack and can be the dominant contribution to the odd-parity transverse voltage. Similar to the USMR measurements, we record ${\Delta R_{\mathrm{yx}} = R_{\mathrm{yx}}(+I_{\mathrm{sense}})-R_{\mathrm{yx}}(-I_{\mathrm{sense}})}$. This ensures that all current-polarity-independent effects are canceled out and only the nonlinear SOT and ANE contributions to the transverse electrical response contribute to the signal\cite{Xue2021}. This method resembles a DC version of the harmonic Hall method, which quantifies these three contributions to determine $B_\mathrm{DL}$ and $B_\mathrm{FL}$. Consequently, we could also have used second harmonic Hall detection to obtain the same signal. For the type XY configuration, we employed the same measurement protocol as for type X switching.

\section{\label{sec:level3}RESULTS AND DISCUSSION\\}
\begin{figure}
\includegraphics[width=\columnwidth]{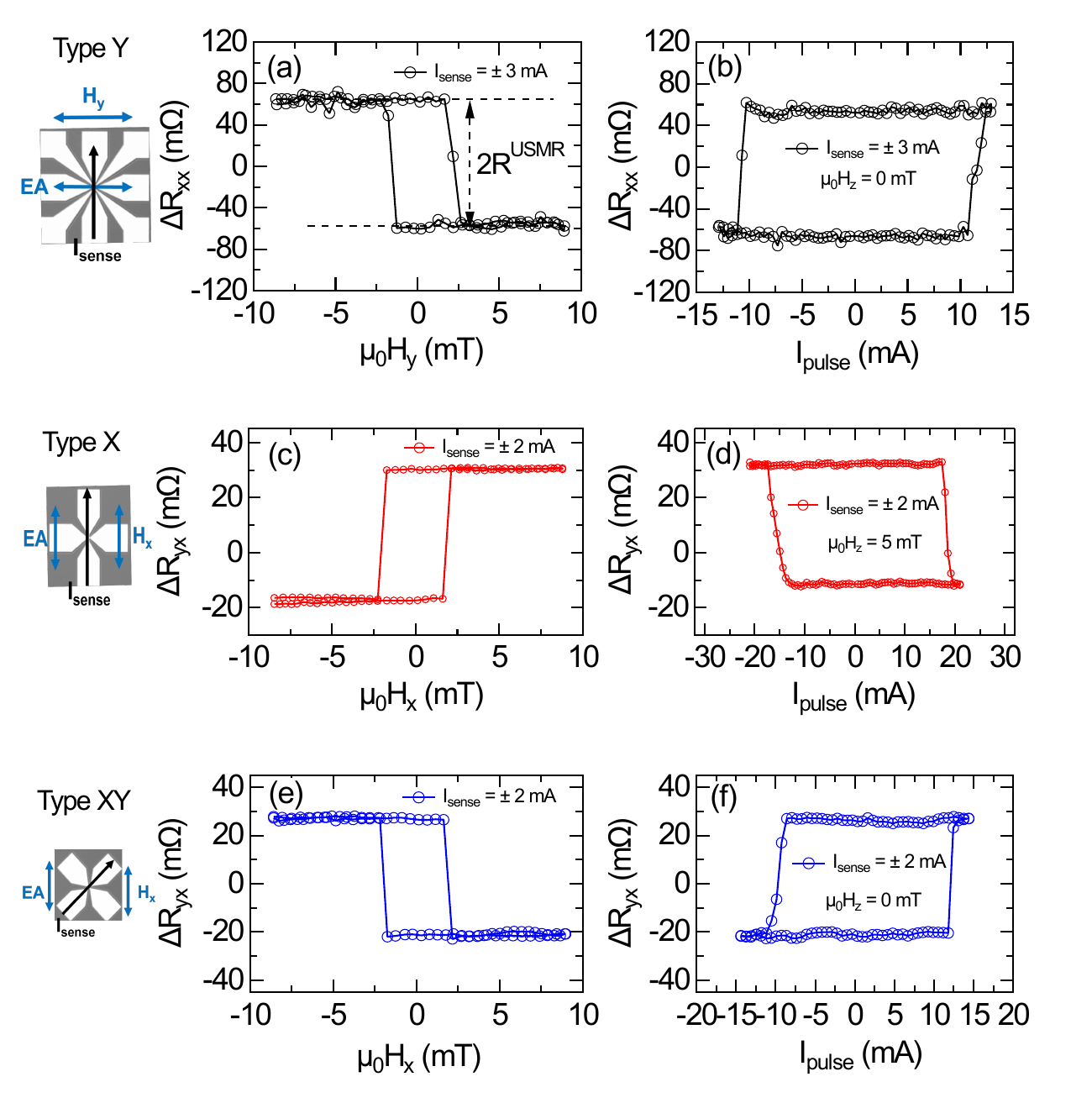}
 \caption{\label{fig : TaCFB2}  Type Y, type X and type XY SOT switching behaviours in Ta/CFB(2)/Pt based on DC USMR and PHE: (a,b) field sweep scan and field-free current-induced switching for type Y (DC USMR), (c,d) Field sweep scan and field ($\mu_0 H_{\mathrm{z}}$) assisted current-induced switching for type X (DC PHE), (e,f) Field sweep scan and field-free current induced switching in type XY (DC PHE).}
\end{figure}

We first demonstrate CIMS measurements in the type Y configuration. Measurements were carried out on sample Ta/CFB(2)/Pt using the USMR technique to probe AMR behavior. An in-plane magnetic field ($\mu_0 H_{\mathrm{y}}$) was swept along the y-direction  while an alternating sense current ($I_{\mathrm{sense}} = \pm\text{3\,mA}, j_{\mathrm{sense}} = \pm1\times10^{11}$\,A/m$^2$) was applied in the x-direction (Fig.1 (d)). The differential longitudinal resistance ($\Delta R_{\mathrm{xx}}$) was recorded throughout. During the $H_{\mathrm{y}}$ sweep, magnetization reversal between $\pm\text{y}$ states gives hysteresis loops with two clear resistance levels as shown in (Fig.~\ref{fig : TaCFB2} (a)). The USMR magnitude is given by $R^{\mathrm{USMR}}=$ $\Delta R_{\mathrm{xx}}/2$ at magnetic saturation (Fig.~\ref{fig : TaCFB2} (a)).
We next measured $\Delta R_{\mathrm{xx}}$ during $\mathrm{~1\,\mu\text{s}}$ width write-pulse current sweeps under identical sense current conditions. Full magnetization reversal between $\pm\text{y}$ states occurred at $\mathrm{+10.7\,mA}$ (L-to-R) and $\mathrm{-10.26\,mA}$ (R-to-L), with an average (arithmetic mean of both polarities) switching current of $\mathrm{10.48\,mA}$ ($j_c = 3.49\times10^{11}$\,A/m$^2$)  (Fig.~\ref{fig : TaCFB2} (b)). A slight asymmetry between the positive and negative switching currents is observed. The asymmetry of switching current is due to asymmetric coercivity that likely arise from domain nucleation imbalance or edge pinning induced during sputtering and fabrication\cite{Xue2023}.
\begin{figure}
\includegraphics[width=\columnwidth]{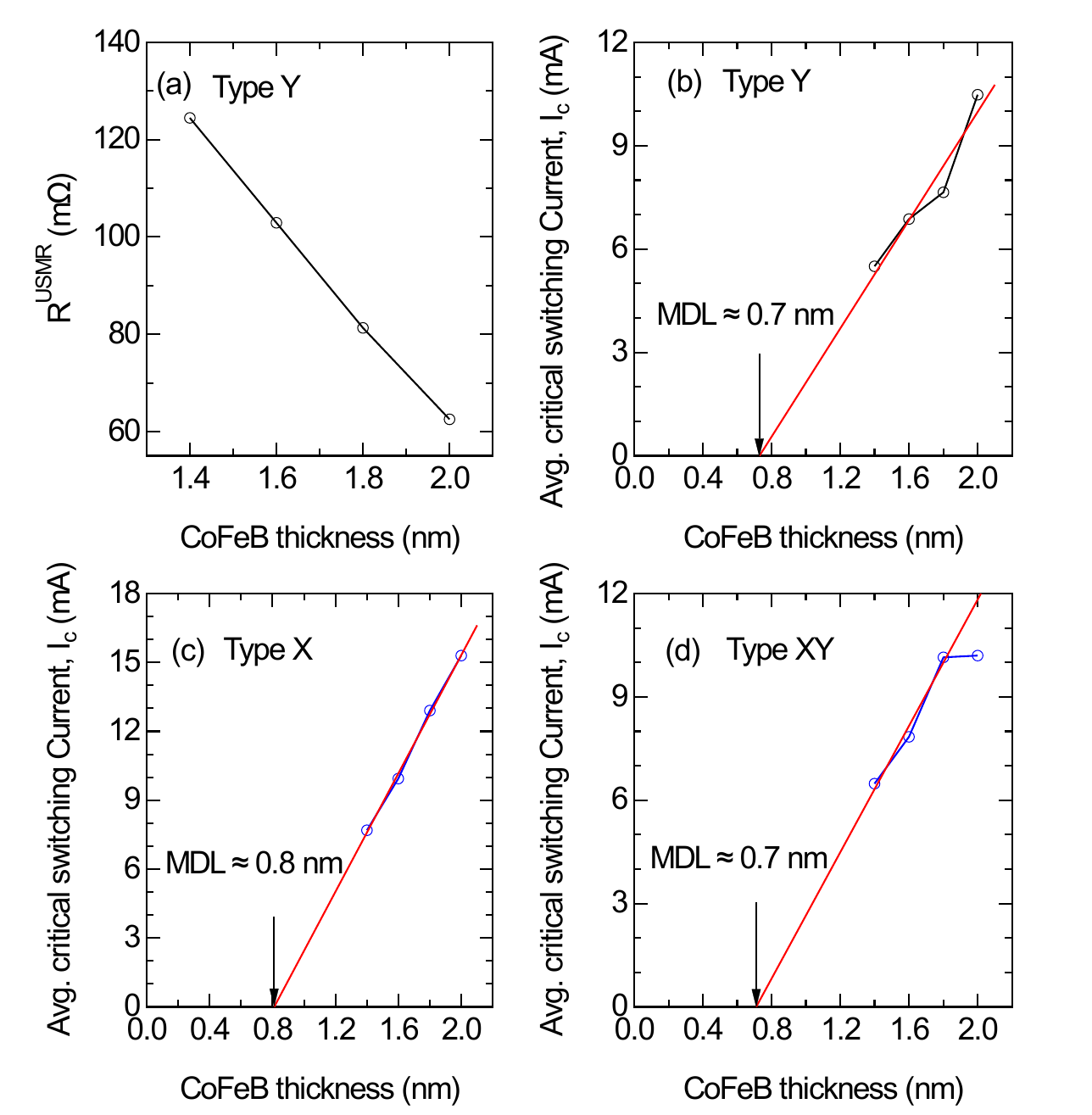}
 \caption{\label{fig : TaCFB_thickness} Type Y, type X and type XY SOT switching behaviours in Ta/CFB(t$_\mathrm{FM}$)/Pt based on DC USMR and  PHE: Thickness dependence of (a) $R^{\mathrm{USMR}}$ for type Y at $I_{\mathrm{sense}}=\pm3\,\text{mA}$, (b) average critical switching current ($I_c$) for type Y at $I_{\mathrm{sense}}=\pm3\,\text{mA}$, (c) $I_c$ for type X at $I_{\mathrm{sense}}=\pm2\,\text{mA}$, (d) $I_c$ for type XY at $I_{\mathrm{sense}}=\pm2\,\text{mA}$, The MDL thicknesses are indicated by the arrows.}
\end{figure}

\begin{figure}
\includegraphics[width=\columnwidth]{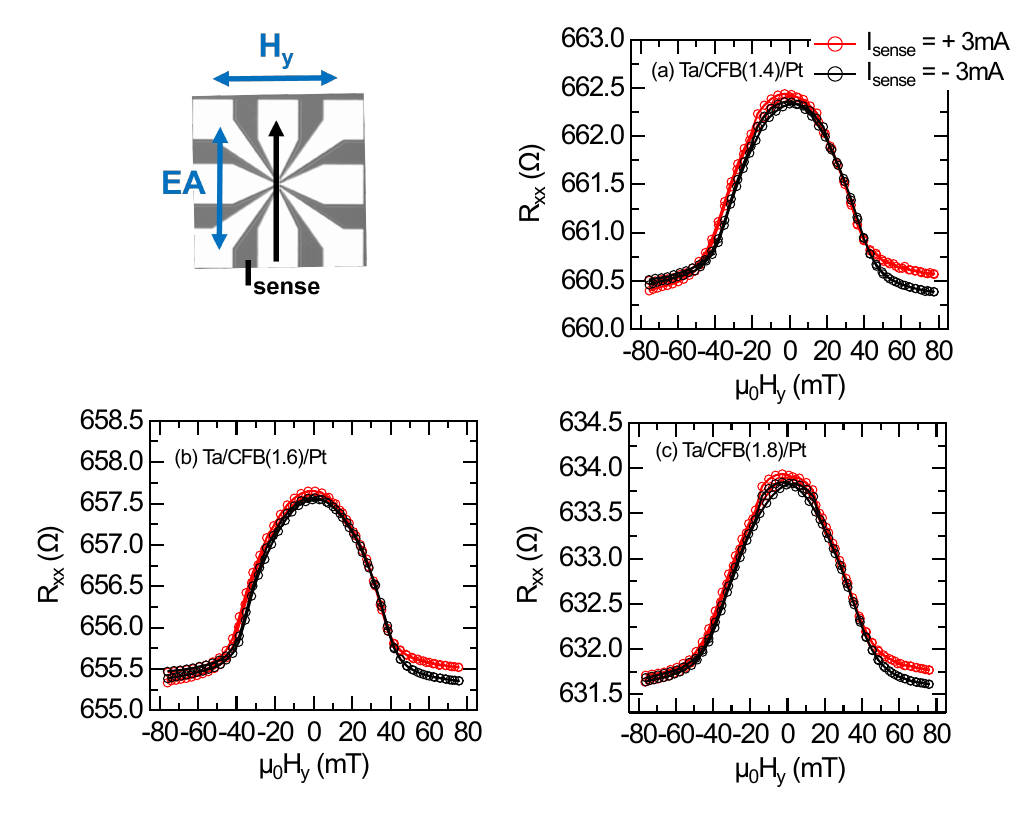}
 \caption{\label{fig : AMR}AMR magnetic hysteresis loops measurement. (a) - (c) Field sweep AMR measurements for device $\mathrm{Ta/CFB(t_{\mathrm{FM}})/Pt}$, $t_{\mathrm{FM}}$ = 1.4, 1.6 and 1.8\,nm by applying the sense current ($I_{\mathrm{sense}} = \pm 3\text{\,mA}$) transverse to the in-plane applied field ($I_{\mathrm{sense}}\parallel \text{EA}$).}
\end{figure}

We now examine type-X and type-XY CIMS, where the EA is parallel or canted to the current. Magnetization detection in these configurations is performed using the DC PHE method. We first verify field-driven magnetization switching in type X devices with a well defined EA, an in-plane field ($\mu_0 H_{\mathrm{x}}$) was swept along the EA under alternating sense current ($I_{\mathrm{sense}} = \pm\text{2\,mA}$, $j_\mathrm{sense} = \pm6.66\times10^{10}$\,A/m$^2$). Figure~\ref{fig : TaCFB2} (c) confirms that the magnetization state in type X devices can be switched and detected reproducibly between the $\pm\text{x}$ orientations.
Next, we examined CIMS using the DC PHE method. We applied $\mathrm{0.4\,\mu s}$ current pulses and achieved full magnetization reversal with the assistance of a $\mathrm{5\,mT}$ out-of-plane field ($\mu_0 H_{\mathrm{z}}$) to break the symmetry, as shown in Fig.~\ref{fig : TaCFB2} (d); consistent with previous reports\cite{Liu2021,Posti2023}, the use of an out-of-plane field represents a state-of-the-art approach for field-assisted Type-X switching. The critical switching currents were $\mathrm{+17.4\,mA}$ (L-to-R) and $\mathrm{-13.2\,mA}$ (R-to-L), with an average of $\mathrm{15.3\,mA}$ ($j_c = 5.1\times10^{11}$\,A/m$^2$), while maintaining the same sensing current used in field-driven switching (Fig.~\ref{fig : TaCFB2} (c)).

In the type XY device, the symmetry is broken intrinsically\cite{Liu2022}, and no external field ($\mu_0 H_{\mathrm{z}}$) is required to achieve complete magnetization switching, as demonstrated in Fig.~\ref{fig : TaCFB2} (e) and (f). In this case, the average critical current is reduced to $\mathrm{10.20\,mA}$ ($j_c = 3.38\times10^{11}$\,A/m$^2$), which is lower than that observed for type X switching in the same sample. This reduction can be attributed to the large canting angle of $45^{\circ}$, which falls into the regime of type Y switching\cite{Liu2022,Liu2021}, where a smaller current density is sufficient to reverse the magnetization. In all three cases, the direct comparison between field and current-induced switching demonstrates the complete magnetization reversal by the CIMS. 
 
The USMR exhibits a decrease with increasing FM thickness, as demonstrated in Fig.~\ref{fig : TaCFB_thickness} (a). Since USMR is predominantly interface sensitive, increasing the FM thickness beyond its spin-diffusion length shunts more current away from the HM/FM interface and reduces interface scattering\cite{Avci2018}. With increasing FM thickness, a linear increase of the switching current is observed (Fig.~\ref{fig : TaCFB_thickness} (b)).
We also observed that the switching current density similarly increases for the type X and type XY configurations (Fig.~\ref{fig : TaCFB_thickness} (c) and (d)), consistent with the decrease of the effective FL and DL fields with increasing FM thickness, as reported by experimental and theoretical studies\cite{Liu2020,Ou2016,manchon2012}.
\begin{figure}
\includegraphics[width=\columnwidth]{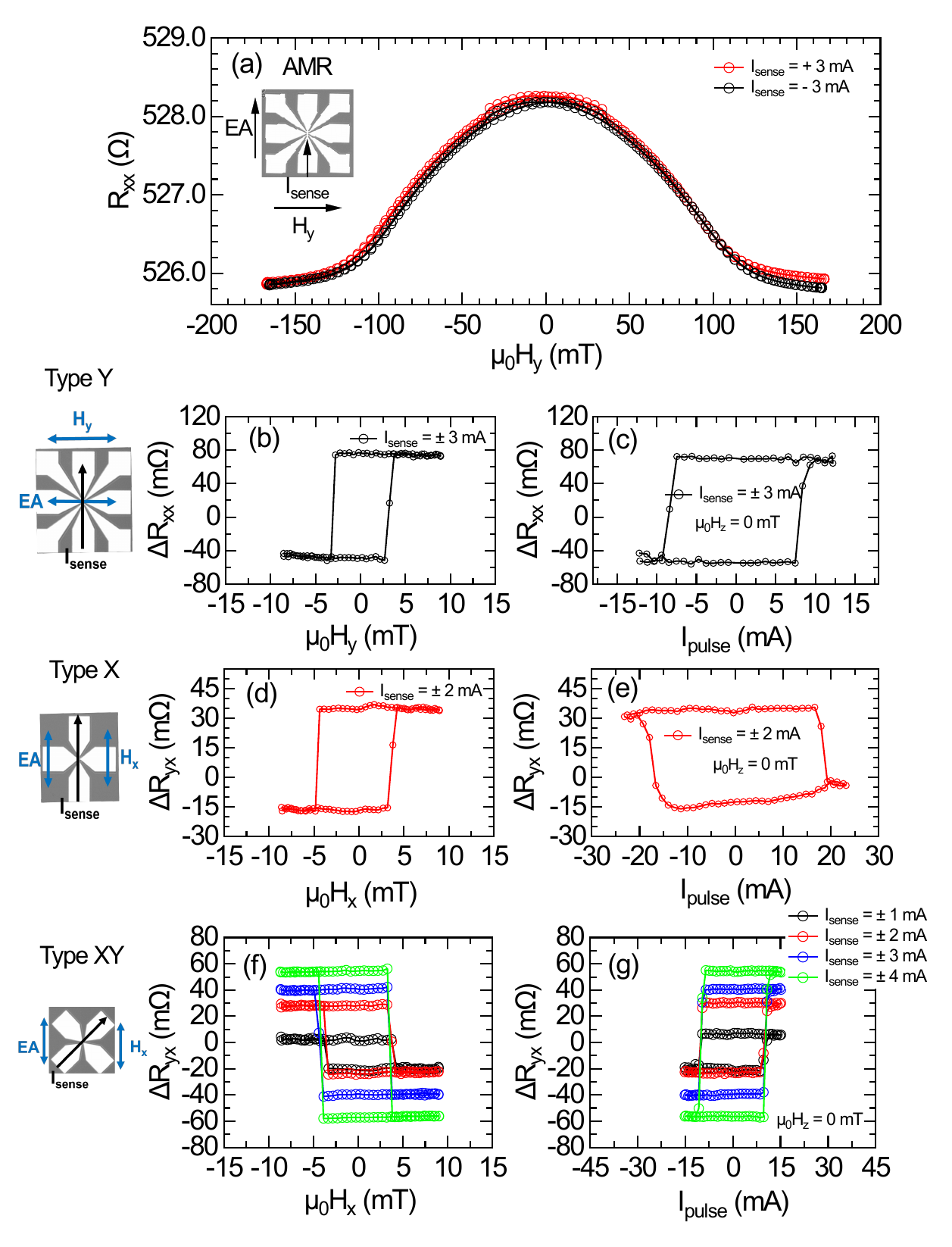}
 \caption{\label{fig : W4CFB1.4} Type Y, type X and type XY SOT switching behaviours in $\mathrm{W (4)/CFB(1.4)/Pt}$ based on USMR and DC PHE: (a) Field sweep AMR measurements by applying the sense current ($I_{\mathrm{sense}}=\pm3\,\text{mA}$) transverse to the in-plane applied field ($I_{\mathrm{sense}}\parallel \text{EA}$), (b,c) field sweep scan and field-free current induced switching for type Y (USMR), (d,e) Field sweep scan and field free current-induced switching for type X (DC PHE), (f,g) Field sweep and field-free current-induced switching in type XY (DC PHE) for different sense ($I_{\mathrm{sense}}$) currents.}
\end{figure}
\begin{figure}
\includegraphics[width=\columnwidth]{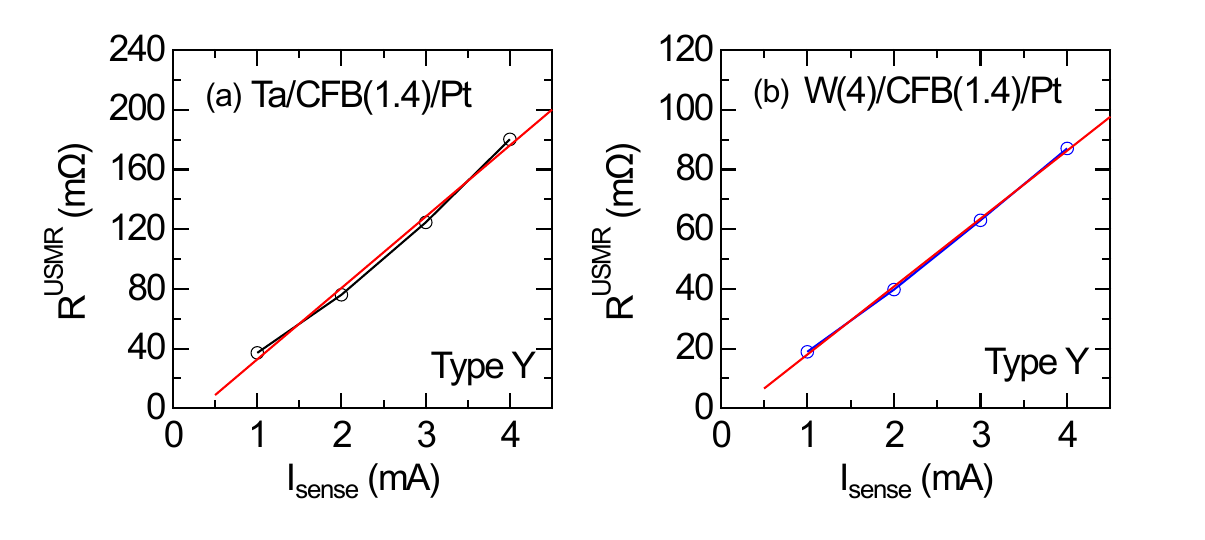}
 \caption{\label{fig : USMR}Sense current ($I_{\mathrm{sense}}$) dependency of USMR ($R^{\mathrm{USMR}}$) for (a) Ta/CFB(1.4)/Pt and (b) W(4)/CFB(1.4)/Pt from the field driven magnetization switching loops.}
\end{figure}
\begin{figure*}
\includegraphics[width=18cm]{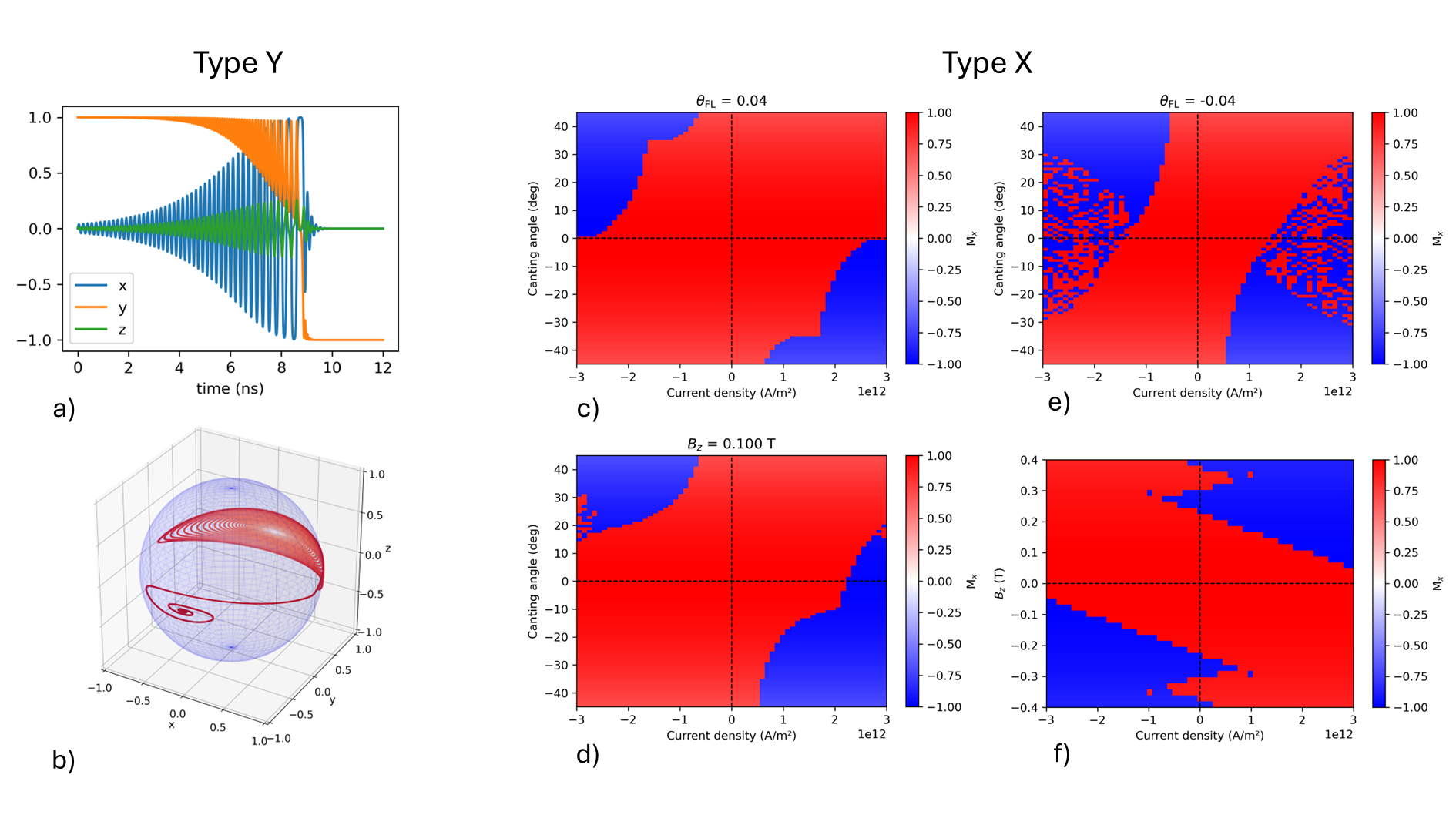}
 \caption{\label{fig:LLG}(a): Cartesian components of the magnetization vector in type Y switching. (b) same as (a), represented on a unit sphere. (c)-(f) type X switching phase diagrams with initial $M_x = +1$. (c) and (d): switching phase diagram with current density, canting angle, for $\theta_\mathrm{FL} = \pm 0.04$. (e) switching phase diagram of current density vs. canting angle for $B_z = 0.1$\,T and $\theta_\mathrm{FL} = 0.0$. (f): switching phase diagram of current density vs. $B_z$ with canting angle of zero and $\theta_\mathrm{FL} = 0.0$. Parameters of the simulations are given in the text.}
\end{figure*}

The linear fits to the $I_c$ vs. CoFeB thickness yield a magnetic dead layer (MDL) of approximately 0.7-0.8\,nm. This value is comparable to MDL values reported for amorphous CoFeB with different underlayers and cap layers\cite{Jang2010}, and is consistent with interfacial magnetic dead layers in Ta/CoFeB-based heterostructures \cite{cecot2017}. The effective FM layer thickness ($t_{\mathrm{FM}}^{\mathrm{eff}} = t_{\mathrm{FM}}-t_{\mathrm{MDL}}$) must be corrected, as it has been reported\cite{Frankowski2015} that in MTJs with PMA, the MDL significantly affects the thermal stability factor and the critical switching current density and plays a significant role in optimizing both the critical current density and thermal stability of the devices.

We did not observe any appreciable change in UMA (${\mu_{\mathrm{0}}H_{\mathrm{a}}\approx\text{60\,mT}}$) for CoFeB thicknesses ranging from 1.4 to 1.8\,nm (Fig.~\ref{fig : AMR} (a)-(c)), except for a slight reduction at 2\,nm, where the $\mu_{0}H_{\mathrm{a}}$ decreases to  approximately 50\,mT (Fig.~\ref{fig : moke} (a)). This trend is expected, as thicker FM layer generally leads to smaller UMA, consistent with the results reported by Fukama et al.\cite{Fukuma2009}.

Further enhancement of UMA can be achieved by depositing a thicker HM underlayer via OAD, as shown here. To investigate the possibility of obtaining stronger torques and stronger anisotropy, we fabricated W(4)/CoFeB(1.4)/Pt sample via OAD and characterized its anisotropy using AMR measurements on micro-sized Hall crosses. An anisotropy field of ${\mu_{\mathrm{0}}H_{\mathrm{a}}\approx\text{146\,mT}}$ was obtained (Fig.~\ref{fig : W4CFB1.4} (a)). The larger anisotropy field is due to the larger W thickness (compared to our Ta samples), which gives a larger ripple amplitude under oblique deposition and thereby enhances the UMA\cite{Scheibler2023}. We then performed CIMS measurements for type Y, type X, and type XY configurations. As shown in Fig.~\ref{fig : W4CFB1.4} (b) and (c), full field and current-driven switching of in-plane magnetization was achieved at a critical switching current ($I_{c} = \pm\text{7.44\,mA}$, $j_c = 2\times10^{11}$\,A/m$^2$) for the type Y configuration. For the type X and type XY configurations, DC PHE measurements confirmed complete in-plane magnetization switching (Fig.~\ref{fig : W4CFB1.4} (d) and (e)). Notably, no external out-of-plane field was applied to break the symmetry. This could arise from a slight misalignment of the current channel with the EA during lithography, as suggested by the small hysteresis jumps observed in the AMR curve (Fig.~\ref{fig : W4CFB1.4} (a)). The field-free switching behavior observed here in the Type X configuration is facilitated by canting angle between the current channel and the EA, a mechanism previously demonstrated to enable deterministic Type X switching\cite{Liu2022}. Although this misalignment occurred accidentally in our device, it is in principle controllable, offering a practical route to achieve robust field-free switching.

For the type XY configuration, field and current-driven switching was studied at different $I_{\mathrm{sense}}$ current (Fig.~\ref{fig : W4CFB1.4} (f) and (g)). No significant influence of $I_{\mathrm{sense}}$ on $I_{c}$ was observed, with the average  $I_{c}$ remaining of 9\,mA. However, a linear dependence of $\Delta R_{\mathrm{yx}}$ on $I_{\mathrm{sense}}$ was found. This is consistent with the expected linear dependence of current-induced SOT field on sense current\cite{Garello2013,Fan2013}. With the thick W layer, both an enhancement of the UMA as well as a reduction of the switching current density $j_\mathrm{c}$ were achieved. The reduced $j_\mathrm{c}$ despite an enhanced UMA can be explained by the enhanced spin Hall angle of W as compared to Ta.

The dependency of USMR ($R^{\mathrm{USMR}}$) on $I_{\mathrm{sense}}$ was studied for the samples Ta/CFB(1.4)/Pt and W(4)/CFB(1.4)/Pt, as shown in Fig.~\ref{fig : USMR} (a) and (b). The USMR increases with increasing $I_{\mathrm{sense}}$ with slopes of $\approx 47\,\mathrm{m\Omega/mA}$ and $23\,\mathrm{m\Omega/mA}$ and demonstrates the expected nonlinear behaviour. The smaller slope in the W(4)/CFB(1.4)/Pt sample is due to the larger shunting of the thicker W layer.
\begin{figure}
\includegraphics[width=\columnwidth]{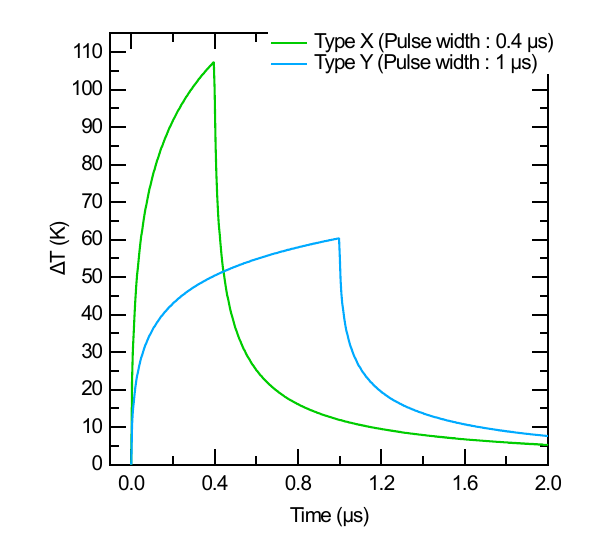}
 \caption{\label{fig : deltaT}Time dependence of the temperature rise $\Delta T$ calculated for type X ($j = 5.1 \times 10^{11}$\,A/m$^2$, pulse width = 0.4\,$\mu$s) and for type Y ($j = 3.49 \times 10^{11}$\,A/m$^2$, pulse width = 1\,$\mu$s).}
\end{figure}
\begin{figure}
\includegraphics[width=\columnwidth]{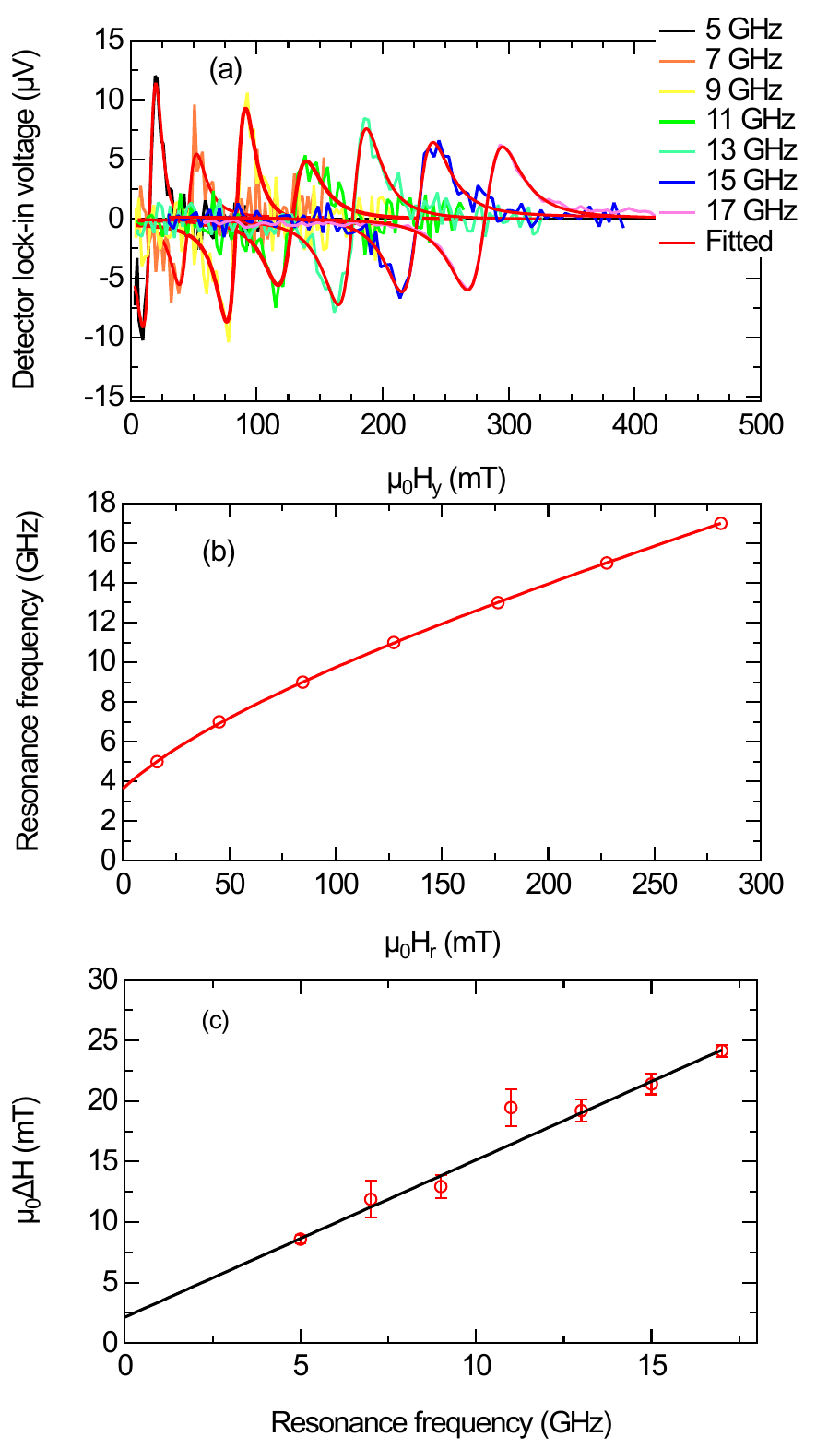}
 \caption{\label{fig : FMR} In-plane FMR measurement of Ta/CFB(2)/Pt: (a) Representative FMR spectra measured at microwave frequencies from 5 to 17\,GHz. (b) Frequency dependence of the resonance field ($\mu_{0}H_{\mathrm{r}}$) and fitted with the Kittel equation (eq.~(\ref{eq:two}))  (c) Frequency dependence of the linewidth fitted using eq.~(\ref{eq:three}). The spectral fits were performed using an asymmetric Lorentzian derivative line shape profile.}
\end{figure}

CIMS measurements were performed on the type Y configuration using pulse widths ranging from $\mathrm{\text{0.4 to 1}\,\mu\text{s}}$. Within this sub-microsecond regime, no thermally activated variations in the critical switching current were observed. In contrast, extended write current pulses in the millisecond to second range require an explicit treatment of thermal activation. The average switching current density ($j_c$) was used to evaluate the DL SOT efficiency\cite{Liu2012} in the macrospin approximation by
\begin{equation}
\xi^{{\mathrm{eff}}}_{\mathrm{DL}} = \frac{2e\mu_{0}M_{\mathrm{s}}t_{\mathrm{FM}}^\mathrm{eff}\alpha_{\mathrm{eff}}(H_{\mathrm{a}}+\frac{M_{\mathrm{eff}}}{2})}{\hbar j_c}
\label{eq:one}
\end{equation}
where $t_{\mathrm{FM}}^\mathrm{eff}$ (MDL corrected effective FM layer thickness), $\alpha_{\mathrm{eff}}$ (effective Gilbert damping parameter), $M_{\mathrm{eff}}$ (effective magnetization), $M_{\mathrm{s}}$ (saturation magnetization). For the heterostructure containing a 2\,nm CoFeB layer, ferromagnetic resonance (FMR) analysis was carried out to obtain $\alpha_{\mathrm{eff}}$, $M_{\mathrm{eff}}$ and  $M_{\mathrm{s}}$ as discussed in the Appendix. The effective SOT efficiency for the 2\,nm CoFeB thickness was found to be 0.20 using eq.~(\ref{eq:one}). This value is slightly lower than reported in the literature using ST-FMR\cite{Huang2018,Skowroski2019} for similar systems, which may result from the lack of MDL correction in previous reports and the FM layer thickness being larger than the characteristic spin dephasing length\cite{Ghosh2012,Sala2022}, both of which can lead to an overestimation of the SOT.

To better understand the observed CIMS, we performed simulations of the Landau-Lifshitz-Gilbert-Slonczewski equation. The following parameters were used: $M_\mathrm{s} = 1050$\,kA/m, $K_\mathrm{u, \perp} = 294$\,kJ/m$^3$ ($B_\mathrm{u, \perp} = 560$\,mT, $M_\mathrm{eff} = 605$\,kA/m), $K_\mathrm{a} = 28.9$\,kJ/m$^3$ ($B_\mathrm{a} = 55$\,mT), $\alpha = 0.039$, $t_{\mathrm{FM}}^\mathrm{eff} = 1.3$\,nm, $\theta_\mathrm{DL} = 0.2$. CIMS was simulated with 10\,ns pulses and a 10\,ns post-pulse equilibration time. A simple solver using the Heun method with a step size of 500\,fs was implemented in \textsc{python} with \textsc{numba}. Results are shown in Fig. \ref{fig:LLG}. For the type Y switching, we obtain deterministic switching with $j = 4.1 \times 10^{11}$\,A/m$^2$, which is very close to the experimental value of Ta/CFB(2)/Pt ($j_c = 3.49\times10^{11}$\,A/m$^2$). The switching trajectory is shown in Fig. \ref{fig:LLG} a) (cartesian components as a function of time) and Fig. \ref{fig:LLG} b) (trajectory on the unit sphere). Type X switching is more nuanced: it explicitly depends on the canting angle between current and EA, the field-like torque component $\theta_\mathrm{FL}$, and the z-axis magnetic field $B_z$. In Fig. \ref{fig:LLG} c) to f) we show switching phase diagrams, starting from $M_x = 1$. Subplots c) and d) demonstrate the effect of the field-like torque for $\theta_\mathrm{FL} = \pm 0.04$, a realistic value that was obtained separately via harmonic Hall measurements on the CFB/Pt interface ($\theta_\mathrm{FL} = + 0.046$). Notably, the Ta/CFB interface has the opposite sign of the FL torque ($\theta_\mathrm{FL} = - 0.036$). As the majority of the current flows through the Pt layer, the CFB/Pt interface is the more relevant one and we thus estimate the total $\theta_\mathrm{FL} \approx + 0.025$, i.e., our calculations slightly overestimate the FL torque. It is clearly seen that the FL torque gives rise to a vastly different switching phase diagram, depending on its sign; only for the negative sign, a random switching regime is observed, whereas the positive sign provides deterministic switching without canting for current densities above $j = 2.5 \times 10^{12}$\,A/m$^2$. With an out-of-plane assist field of $B_z = 100$\,mT, deterministic switching without canting is obtained at $j = 2.1 \times 10^{12}$\,A/m$^2$. In Figure \ref{fig:LLG} f) we show the dependence of the switching current density on $B_z$ and observe that $j_\mathrm{c}$ depends linearly on $B_z$. For large $B_z$, a toggle regime around zero current density is observed.

To summarize, our observed switching current density of the type X configuration is not explained by the LLGS simulations, which indicate significantly higher $j_\mathrm{c}$ with any realistic set of parameters. Thus, we conclude that the switching mechanism is of the nucleation and domain wall propagation type, rather than macrospin switching. In contrast, the type Y switching can be explained as macrospin switching. Type XY (at 45$^\circ$) has $j_\mathrm{c} \approx 5 \times 10^{11}$\,A/m$^2$, close to our experimental findings and independent of $B_z$ and $\theta_\mathrm{FL}$; it thus clearly resembles the type Y switching, which shares these features with type XY (not shown). 
As a final note, we mention that the effect of temperature is not considered in these calculations. We next consider the influence of Joule heating in the present pulse width ($\tau_{\mathrm{pulse}}$) regime. The temperature rise at end of the pulse, $\Delta T$, was estimated using a two-dimensional heat-diffusion model\cite{you2006,Fritz2020}. The model equation explicitly treats a metal layer on a homogeneous substrate. In our devices, a 200\,nm $\mathrm{SiO_2}$ layer is present, which modifies the heat flow and therefore the effective temperature scaling. To account for this effect, we performed stationary finite-element (FEM) simulations of the heat transport including the $\mathrm{SiO_2}$ layer. From these simulations, we find that the model result must be multiplied by a thermal scaling factor (c = 2.85). Using this scaling, the estimated temperature rise is $\Delta T \approx 60.33$\,K for type Y ($j = 3.49 \times 10^{11}$\,A/m$^2$, $\tau_{\mathrm{pulse}}$ = 1\,$\mu$s) and $\Delta T \approx 107.37$\,K for type X ($j = 5.1 \times 10^{11}$\,A/m$^2$, $\tau_{\mathrm{pulse}}$ = 0.4\,$\mu$s) (Fig.~\ref{fig : deltaT}). The corresponding reduction of the critical switching current density can be obtained by using an Arrhenius-type relation\cite{Fritz2020,Liu2021} 
\begin{equation}
\frac{j_\mathrm{c}}{j_\mathrm{{c,0}}} = 1- \frac{1}{\Delta}\ln\!\left(\frac{\tau_{\mathrm{pulse}}}{\tau}\right)
\label{eq:two}
\end{equation}
where $\Delta = E_\mathrm{b} / k_\mathrm{B} (T + \Delta T)$ is the thermal stability factor, $E_\mathrm{b}$ is the energy barrier, $k_\mathrm{B}$ is Boltzmann's constant, $T = 293\,\mathrm{K}$ is room temperature, and $\tau = 1\,\mathrm{ns}$ is the attempt timescale. Since $E_\mathrm{b}$ cannot be directly extracted from our experiments, representative literature values of $E_\mathrm{b} = 1.0$ and $1.5\,\mathrm{eV}$ were adopted, corresponding to thermal stability factors of approximately 29 and 49, respectively\cite{Takeuchi2015,Guo2023,Liu2021}. The resulting ratios $j_\mathrm{c}/j_\mathrm{{c,0}} \approx 0.79$-$0.88$ for all switching geometries indicate that Joule heating may reduce $j_\mathrm{c}$ by 12-21\,\%.
\section{\label{sec:level3}CONCLUSION\\}
In summary, we used DC USMR and PHE to study sub-microsecond CIMS in in-plane magnetized trilayer systems with easy-axis defined by oblique angle deposition. In (Ta\ or\ W)/CoFeB/Pt structures, the CoFeB layer sandwiched between two HMs with opposite SHA shows large Gilbert damping and large SOT efficiency. The reduction in the critical switching current density in Ta/CFB(1.4)/Pt by $45\,\%$ compared to Ta/CFB(2)/Pt is accompanied by an increase in USMR with decreasing CoFeB thickness. The electrical detection via direct current USMR or PHE was shown and complete magnetization reversal for type X and type Y configurations was demonstrated. For a quantitative understanding, the magnetic dead layer was properly included in the analysis and was easily obtained from the switching current density measurements with varying FM thickness. The type Y switching can be directly explained within a macrospin model; the type X switching, however, is more nuanced and most likely progresses via domain nucleation and propagation, which reduces the critical current density compared to the macrospin model. These results highlight that combining oblique angle deposition of the heavy-metal underlayer with easy-axis engineering for field-free type X and type Y switching yields efficient spintronic devices for future low-power, high-density, and fast memory applications.
\begin{acknowledgments}
We acknowledge the financial support by the Deutsche Forschungsgemeinschaft under Project No. 513154775 and No. 518575758, and by the DFG Major Research Instrumentation programme Project No. 511340083 and Project No. 468939474.  We gratefully acknowledge Prof. Lambert Alff at Technical University Darmstadt for providing us with access to his  laboratory's X-ray diffractometer.
\end{acknowledgments}

\appendix
\section{}

In-plane FMR measurements were carried out on the sample Ta (2\,nm)/CoFeB (2\,nm)/Pt (2\,nm), over the frequency range of 5-17\,GHz in the field-sweep mode to investigate the magnetization relaxation (Fig.~\ref{fig : FMR} (a)). We used the OpenFMR measurement system\cite{Tdos2025}. The FMR spectra were analyzed using the derivative of an asymmetric Lorentzian line-shape function, from which the resonance field ($\mu_{0}H_{\mathrm{r}}$) and linewidth ($\mu_{0}\Delta H$) were extracted (Fig.~\ref{fig : FMR} (b) and (c)]. The frequency dependence of the resonance field was fitted with the Kittel relation(eq.~(\ref{eq:two})), while the linewidth variation with frequency was fitted using (eq.~(\ref{eq:three}))\cite{hait2022,Tdos2025}. The effective Gilbert damping parameter ($\alpha_{\mathrm{eff}}$) and the effective magnetization ($M_{\mathrm{eff}}$), for the Ta (2\,nm)/CoFeB (2\,nm)/Pt (2\,nm) heterostructure, are $\alpha_{\mathrm{eff}} = 0.039\pm0.001$ and $M_{\mathrm{eff}} = 608.6\pm1.256\,\text{kA/m}$ and the saturation magnetization ($M_{\mathrm{s}}$) was estimated to be approximately $1050\,\text{kA/m}$, as measured with a 20nm thick CoFeB layer.
\begin{equation}
f = \frac{\mu_{0}\gamma}{2\pi}\sqrt{(H_{\mathrm{r}}+H_{\mathrm{a}})(H_{\mathrm{r}}+H_{\mathrm{a}}+M_{\mathrm{eff}})}
\label{eq:two}
\end{equation}
\begin{equation}
\mu_0\Delta H = \mu_0\Delta H_0+\frac{4\pi\alpha_{\mathrm{eff}}}{\gamma}f
\label{eq:three}
\end{equation}

\bibliography{CIMS_library1}

\end{document}